\documentstyle[12pt,a4]{article}

\begin{document}

\title{\sc SOLUTIONS OF THE
EINSTEIN-MAXWELL-DIRAC AND SEIBERG-WITTEN MONOPOLE EQUATIONS }
\author{Cihan Sa\c{c}l\i o\~{g}lu$^{1,2}$\\
\date{$^{1}$Physics Department, Bo\~{g}azi\c{c}i University \\
80815 Bebek--\.{I}stanbul, Turkey\\
and \\
$^{2}$Feza G\"{u}rsey Institute \\   TUBITAK--Bo\~{g}azi\c{c}i University\\ 
81220 \c{C}engelk\"{o}y, {I}stanbul-- Turkey}}
\maketitle
%\vspace*{1cm}
\begin{abstract}
We present unique solutions of the Seiberg-Witten Monopole Equations
in which the $U(1)$  curvature is covariantly constant, the monopole
Weyl spinor consists of a single constant component, and the 4-manifold
is a product of two Riemann surfaces of genuses $p_{1}$ and $p_{2}$.
There are $p_{1} -1$ magnetic vortices on one surface and $p_{2} - 1$
electric ones on the other,
with $p_{1} + p_{2} \geq 2$ ($p_{1} = p_{2}= 1$ being excluded).
When $p_{1} = p_{2}$, the electromagnetic fields are self-dual and
one also has a solution of the coupled euclidean Einstein-Maxwell-Dirac
equations, with the monopole condensate serving as cosmological
constant.  The metric is decomposable and the electromagnetic fields
are covariantly constant as in the Bertotti-Robinson solution.
The Einstein metric can also be derived from a K\"{a}hler
potential satisfying the Monge-Amp\`{e}re equations.
\end{abstract}
\vspace*{3 cm}
\pagebreak
\baselineskip=30pt

\noindent  {\bf 1.  Introduction:}

Four-dimensional manifolds of Euclidean signature are of interest
in physics for semiclassical treatments of vacuum tunnelling and
for their contributions to the path integral of Quantum Gravity.
In such a physical context,  one might start by trying to solve
Einstein's equations in the presence of some chosen set of  matter
sources, together with field equations (adapted to the background metric)
for the matter fields.  In mathematics, on the other hand, four
dimensions are unique in hosting infinitely many manifolds that
are homeomorphic but  not diffeomorphic to each other.  The most
efficient approach for classifying such manifolds involves the
Seiberg-Witten monopole equations (SWME)\cite{Wit1}
\begin{equation}\label{1}
\not{\! \! D}_{A} \psi = 0,
\end{equation}
\begin{equation}\label{2}
F^{+}_{\mu\nu}\equiv\frac{1}{2}(F_{\mu\nu} + 
\frac{1}{2}\epsilon_{\mu \nu \alpha \beta}F^{\mu\nu})
=-\frac{i}{4} \psi^{\dag}[\gamma_{\mu},\gamma_{\nu}]\psi~~~.
\end{equation}
\noindent In (\ref{1}) and (\ref{2}) $A_{\mu}$ is the $U(1)$
connection and $F^{+}_{\mu\nu}$ is the self-dual part
of $F_{\mu\nu}=\partial_{\mu}A_{\nu}-\partial_{\nu}A_{\mu}$;
the covariant Dirac operator also involves the spin connection of the
4-manifold ${\cal M}_{4}$.  For a pedagogical review of  the SWME,
we refer the reader to \cite{Akb}.  The equations stem from
an $N=2$ supersymmetric Yang-Mills theory which is first "twisted"
into a topological  quantum field theory (TQFT) \cite{Wit2} and
then has its gauge symmetry broken down spontaneously from
$SU(2)$ to $U(1)$.  For the particular choice of Higgs vacuum
used in the SWME, the classical monopole solutions, represented
by the Weyl spinor $\psi$, become massless, which is why a mass
term is absent in (\ref{1}).

In the mathematical study of 4-manifolds, the focus is usually
on global questions  (e.g. topological invariants, the moduli
space  of a class of solutions etc.) rather than explicit local
solutions of the SWME.  However, it is instructive to derive
from the SWME, in actual local form, a solution whose global
properties are known.  Another reason for searching for local
solutions is the possibility  that some of these may represent
configurations of physical significance, which may then indicate
new connections between the topology of 4-manifolds and physics.
In particular, it is important to note that there are {\em two}
SWME equations for {\em three} fields; the fact that  the metric
is not constrained by a third equation is of course consistent
with the topological nature of the SW system.  If one were to
consider a "physical" equation for the metric, the obvious
candidate would of course be Einstein's field equations in the
presence of a Maxwell and a Dirac (or, rather, Weyl) field.
To make the system completely physical, Maxwell's equations would also have to
be added, resulting in an overdetermined system because of (\ref{2}).
LeBrun \cite{LeBrun} has in fact recently considered manifolds
simultaneously obeying the SWME and Einstein's equations with a
cosmological term {\em without} the energy-momentum tensors of
the Maxwell and Dirac fields on the RHS; we will see in a specific
example below how and when this can be justified.

Remarkably, a set of solutions to this overconstrained system exist
as reported briefly in \cite{Sac}.  One of the purposes of the
present paper is to exhibit their derivation in detail; but we
start with our more general SWME solutions of which the "physical"
ones are a  special subset:  The 4-manifold ${\cal M}_{4}$ is a
product of two Riemann surfaces
$\Sigma_{p_{1}}$ and  $\Sigma_{p_{2}}$, where the genuses $p_{1}$
and $p_{2}$ must satisfy
$p_{1}+p_{2}\geq 2$, excluding $p_{1}=p_{2}=1$.  
The Weyl spinor $\psi$ consists of a single constant component
$\psi_{1}$ or $\psi_{2}$, which may be interpreted as a monopole
condensate.  It is worth remarking that in the closely related
"physical" Seiberg-Witten theory \cite{Wit3} based on an untwisted
$N=2$ supersymmetric Yang-Mills theory in flat Minkowski
space, precisely such a condensate leads to quark
confinement, while   a "gluino condensate", possibly dual
to the monopole one, is considered in \cite{Wit2}.
Physically, $p_{1} -1~(p_{2}-1)$ is the number
of magnetic (electric) vortices in $\Sigma_{p_{1}}~( \Sigma_{p_{2}})$.  

Interestingly, it is the most symmetric special
case (with self-dual electromagnetic fields) of the
above SWME solutions  that solves the coupled
Einstein-Maxwell-Dirac equations, with the condensate
now serving as the cosmological constant.  The solution is
reminiscent  of the Bertotti-Robinson one \cite{B3}, \cite{R4}
in that the metric is decomposable and the electromagnetic
fields are covariantly constant.  However, the Euclidean
signature and the spinor ({\it not}  the cosmological
constant,  which is present in \cite{B3}) are new features.
The presence of the spinor in particular provides a
counterexample to the folk-theorem that one need not seek
solutions of the coupled Einstein-Maxwell-Dirac equations
since the Dirac field supposedly becomes negligible in the
classical limit where the field  equations apply.

The plan of the paper is as follows.  In section 2, we describe
the Ansatz for the SWME, which are then reduced to a pair of
Liouville equations. In section 3, we discuss the behavior
of the Ansatz under gauge transformations.  The solutions, which
are expressed in terms of Fuchsian functions, are classified
and presented in section 4.  We compute the topological numbers
of the solutions in section 5, and using these, prove their
uniqueness by showing that the virtual dimension of the moduli
space of solutions is zero.  In section 6, we show that the
special self-dual SWME solution is the only one which also
satisfies the Einstein-Maxwell-Dirac equations.  In
section 7, the K\"{a}hler form of the metric and the
Monge-Amp\`{e}re equation satisfied by the K\"{a}hler \
potential are given. Section 8 ends the paper with
concluding remarks.

\noindent {\bf 2.   An  Ansatz based on
${\cal M}^{(1)}_{2}\times{\cal M}^{(2)}_{2}$  :}

A central tool in Seiberg-Witten  theory of 4-manifolds is the
Weitzenbock formula obtained by squaring  $\not{\! \! D}_{A}$
in (\ref{1}); together with (\ref{2}), it can be shown to imply
that 4-manifolds with everywhere non-negative scalar curvatures
cannot admit non-singular solutions.  Integrating the same
formula, Witten derived "vanishing theorems" showing that
the solutions for non-negative curvature are not only
singular, but also non-square integrable.  Thus we expect that
at least one of the two Riemann surfaces will have
genus $p \geq 2$; if one surface is the two-sphere, the
curvature of the higher genus surface must be sufficiently
large and negative to make the overall scalar curvature
negative.  This expectation will be explicitly verified by
the computation that follows.

We start with the conformally flat basis 1-forms 
\begin{equation}\label{3}
e^{i}=e^{\mu}dx^{i},  \mu=\mu(x^{1},x^{2}), i=1, 2;~~ e^{j}=e^{\nu}dx^{j},  
\nu=\nu(x^{3},x^{4}), j=3,4 
\end{equation}
\noindent for ${\cal M}^{(1)}_{2}$ and ${\cal M}^{(2)}_{2}$,
respectively.  We also choose the similarly "decomposable" $U(1)$
connection
\begin{equation}\label{4}
A_{\mu}=(A_{1}(x^{1},x^{2}),
A_{2}(x^{1},x^{2}),A_{3}(x^{3},x^{4}),A_{4}(x^{3},x^{4})),
\end{equation}
\noindent while the spinor $\psi^{T}=
(\psi_{1}, \psi_{2}, 0, 0)$ is initially assumed to depend on
all four coordinates.  The  Cartan structure equations yield
the spin-connection 1-forms
\begin{equation}\label{5} 
\omega^{12}= \partial_{2} \mu ~ dx^{1}- \partial_{1} \mu  ~ dx^{2},~ 
\omega^{34}= \partial_{4} \nu ~ dx^{3}- \partial_{3} \nu ~ dx^{4}
\end{equation}
\noindent  with all other components vanishing.  Of the curvature 2-forms
\begin{equation}\label{6} 
R^{a}_{b} = d\omega^{a}_{b}+\omega^{a}_{c} \wedge \omega^{c}_{b}
\end{equation}
\noindent the only non-zero ones then are
\begin{equation}\label{7} 
R^{1}_{2} = -(\partial_{1}\partial_{1}+\partial_{2}\partial_{2})\mu
~ dx^{1}\wedge dx^{2}=
- [(\partial_{1}\partial_{1}+\partial_{2}\partial_{2})] e^{-2\mu} e^{1}
\wedge  e^{2}
\end{equation}
\noindent and
\begin{equation}\label{8} 
R^{3}_{4} = -(\partial_{3}\partial_{3}+\partial_{4}
\partial_{4})\nu ~ dx^{3} \wedge dx^{4}=
- [(\partial_{3}\partial_{3}+
\partial_{4}\partial_{4})] e^{-2\nu} e^{3} \wedge  e^{4}.
\end{equation}
\noindent We define the curved-index $\gamma$-matrices via
\begin{equation}\label{9} 
\gamma^{\mu}=\gamma^{a} E^{\mu}_{a},
\end{equation}
\noindent where $E^{\mu}_{a}e^{b}_{\mu}=\delta^{a}_{b}$.  Hence
\begin{equation}\label{10} 
E_{1}=e^{-\mu}\partial_{1}, ~ E_{2}=e^{-\mu}\partial_{2}, ~
E_{3}=e^{-\nu}\partial_{3}, ~ E_{4}=e^{-\nu}\partial_{4}.
\end{equation}
\noindent We use the flat space $\gamma$-matrices
\begin{eqnarray}\label{11}
\gamma^1 &=& \tau_1 \otimes \sigma_1 = 
             \left( \begin{array}{cc} {\bf 0} & {\bf \sigma}_1 \\
             {\bf \sigma}_1 & {\bf 0} \end{array}\right)~~,~~
           \gamma^2 = \tau_1 \otimes \sigma_2 = 
           \left( \begin{array}{cc} {\bf 0} & {\bf \sigma}_2 \\
           {\bf \sigma}_2 & {\bf 0} \end{array}\right)~~, \nonumber \\
\gamma^3 &=& \tau_1 \otimes \sigma_3 =
             \left( \begin{array}{cc} {\bf 0} & {\bf \sigma}_3 \\
             {\bf \sigma}_3  & {\bf 0} \end{array}\right)~~,~~
           \gamma^4 = \tau_2 \otimes 1 = 
           \left( \begin{array}{cc} {\bf 0} & -i1 \\
           i1 & {\bf 0} \end{array}\right)~~, \nonumber \\
\gamma^5 &\equiv& 
         \gamma^4 \gamma^1 \gamma^2 \gamma^3 = \left( \begin{array}{cc}
              1 & {\bf 0} \\
              {\bf 0} & -1 \end{array}\right)~~.
\end{eqnarray}
\noindent The explicit form of the Dirac equation thus  becomes
\begin{equation}\label{12}
\not{\! \! D}_{A} \psi \equiv \gamma^{a} E^{\mu}_{a}
(\partial_{\mu}+iA_{\mu}+\frac{1}{8}
\omega^{bc}_{\mu}[\gamma_{b},\gamma_{c]})\psi=0,
\end{equation}
\noindent  while (\ref{2}) reduces to the three equations
\begin{equation}\label{13}
2F^{+}_{12}= \partial_{1}A_{2}  - \partial_{2}A_{1} +
(\partial_{3}A_{4}  - \partial_{4}A_{3}) e^{2(\mu-\nu)} =
e^{2\mu}( |\psi_{1}|^{2} - |\psi_{2}|^{2}),
\end{equation}
\begin{equation}\label{14}
\bar{\psi}_{1}\psi_{2}+\bar{\psi}_{2}\psi_{1} = i(\bar{\psi}_{2}\psi_{1}  -
\bar{\psi}_{1}\psi_{2})=0
\end{equation}
\noindent since $F_{13}=F_{24}=F_{23}=F_{14}=0$.  In component
form, (\ref{12}) reads
\begin{equation}\label{15}
e^{-\nu}(\partial_{3}+iA_{3}+i\partial_{4}-A_{4}+{1 \over  2}
\partial_{3}\nu +{i  \over 2}\partial_{4}\nu)\psi_{1}+e^{-\mu}
(\partial_{1}+iA_{1}-i\partial_{2}+A_{2}+{1 \over  2}
\partial_{1}\mu - {i  \over 2}\partial_{2}\mu)\psi_{2}=0,
\end{equation}
\begin{equation}\label{16}
e^{-\mu}(\partial_{1}+iA_{1}+i\partial_{2}-A_{2}+{1 \over  2}
\partial_{1}\mu +{i  \over 2}\partial_{2}\mu)\psi_{1}+
e^{-\nu}(-\partial_{3}-iA_{3}+i\partial_{4}-A_{4}-
{1 \over  2} \partial_{3}\nu + {i  \over 2}\partial_{4}\nu)\psi_{2}=0.
\end{equation}
\noindent Equation (\ref{14}) shows we must set either
$\psi_{1}$ or $\psi_{2}$ equal to zero.  Let us start
with $\psi_{2}=0$.  Then  (\ref{13}) becomes
\begin{equation}\label{17}
(\partial_{1}A_{2}  - \partial_{2}A_{1})e^{-2\mu} +
(\partial_{3}A_{4}  - \partial_{4}A_{3}) e^{-2\nu)}=|\psi_{1}|^{2}.
\end{equation}
\noindent Now, the first term in (\ref{17}) is a function of 
$(x^{1}, x^{2})$ and the second of $(x^{3}, x^{4})$.  The
simplest choice for the RHS compatible with this splitting is
\begin{equation}\label{18}
\psi_{1}=\zeta (x^{1}, x^{2})+i \eta (x^{3}, x^{4}),
\end{equation}
\noindent where $\zeta$ and $\eta$ are real functions.
Putting (\ref{18}) in (\ref{15}) and (\ref{16}) then gives
\begin{equation}\label{19}
[iA_{3} - A_{4}+({1 \over  2} \partial_{3}\nu +
{i  \over 2}\partial_{4}\nu)]\zeta =0,
\end{equation}
\noindent or
\begin{equation}\label{20}
A_{3} = - {1 \over  2} \partial_{4}~\nu,~~~A_{4} =
{1 \over  2} \partial_{3}~\nu ,
\end{equation}
\noindent and hence 
\begin{equation}\label{21}
(\partial_{3}+ i \partial_{4}) \eta = 0. 
\end{equation}
\noindent  Now since $\eta$ is real by (\ref{18}), (\ref{21})
does not mean $\eta$ is a holomorphic function; in fact, it can
only be a real constant.  Using  (\ref{20}), we then have
\begin{equation}\label{22}
F_{34} = {1 \over  2} (\partial_{3}\partial_{3}+
\partial_{4} \partial_{4})\nu.
\end{equation}
\noindent A similar analysis of (\ref{16}) gives
\begin{equation}\label{23}
A_{1} = - {1 \over  2} \partial_{4}~\mu,~~~A_{2} =
{1 \over  2} \partial_{3}~\mu ,
\end{equation}
\begin{equation}\label{24}
(\partial_{1}+ i \partial_{2}) \zeta = 0,
\end{equation}
\begin{equation}\label{25}
F_{12} = {1 \over  2} (\partial_{1}\partial_{1}+
\partial_{2} \partial_{2})\mu.
\end{equation}
\noindent Thus $\zeta$ is also a real constant. Putting these
back in (\ref{17}), we find
\begin{equation}\label{26}
e^{-2\mu}(\partial_{1}\partial_{1}+
\partial_{2} \partial_{2})\mu + e^{-2\nu}(\partial_{3}\partial_{3}+
\partial_{4} \partial_{4})\nu =  2({\eta}^{2}+{\zeta}^{2}),
\end{equation}
\noindent or, comparing with (\ref{7}) and (\ref{8}), 
\begin{equation}\label{27}
R=2R^{12}_{12}+2R^{34}_{34}= - 2({\eta}^{2}+{\zeta}^{2})=
- 2|\psi_{1}|^{2}.
\end{equation}
\noindent We thus see that the Ricci curvature $R$ of the
manifold ${\cal M}_{4}$ is a negative constant and non-singular
solutions are allowed.  We can satisfy (\ref{26}) and (\ref{27})
by setting
\begin{equation}\label{28}
e^{-2\mu}(\partial_{1}\partial_{1}+
\partial_{2} \partial_{2})\mu = 2 |\phi|^{2}= - 2R^{12}_{12} 
\end{equation}
\begin{equation}\label{29}
e^{-2\nu}(\partial_{1}\partial_{1}+
\partial_{2} \partial_{2})\nu = 2 (|\psi_{1}|^{2} - |\phi|^{2})
= - 2R^{34}_{34},
\end{equation}
\noindent where we have introduced another constant $|\phi|$.
The options for
${\cal M}^{(1)}_{2}$ and ${\cal M}^{(2)}_{2}$ are now related
to the values of $|\phi|$ and $|\psi_{1}|$ in the following way:

(i) Both manifolds have constant negative curvature if
$|\psi_{1}|>|\phi| \neq 0$.

(ii) ${\cal M}^{(1)}_{2}$ has zero and ${\cal M}^{(2)}_{2}$
has negative curvature if $|\phi|=0$.

(iii) ${\cal M}^{(2)}_{2}$ has zero and ${\cal M}^{(1)}_{2}$
has negative curvature if $|\phi|=|\psi_{1}|$.

(iv) ${\cal M}^{(1)}_{2}$ has constant negative and 
${\cal M}^{(2)}_{2}$ has constant positive curvature if
$|\psi_{1}|<|\phi| $.

At this point, it is advantageous to go over to the two pairs
of dimensionless coordinates
$(x,y) \equiv \sqrt{2} |\phi| (x^{1},x^{2})$ and $(s,t)
\equiv \sqrt{2} (|\psi_{1}|^{2} -
|\phi|^{2})^{1/2} (x^{3},x^{4})$ and then to define their
complex combinations
\begin{equation}\label{30}
z^{1}\equiv x+iy,~~~z^{2}\equiv s+it.
\end{equation}
\noindent One can now rewrite (\ref{28}) and (\ref{29}) as the
pair of Liouville equations
\begin{equation}\label{31}
4\partial_{1} \partial_{\overline{1}}\mu = e^{2\mu}, 
\end{equation}
\begin{equation}\label{32}
4\partial_{2} \partial_{\overline{2}}\nu = \pm e^{2\nu}~~(|\psi_{1}|
\stackrel{>}{<} |\phi|). 
\end{equation}
\noindent Cases (i) and (iv) are represented by equation (\ref{31})
combined with the upper and lower  sign versions of
(\ref{32}), respectively.  For case (ii) we have the pair
of equations
\begin{equation}\label{33}
\partial_{1} \partial_{\overline{1}}\mu = 0,
\end{equation}
\begin{equation}\label{34}
4\partial_{2} \partial_{\overline{2}}\nu = e^{2\nu}, 
\end{equation}
\noindent while case (iii) is described  by the pair
\begin{equation}\label{35}
4\partial_{1} \partial_{\overline{1}}\mu = e^{2\mu} ,
\end{equation}
\begin{equation}\label{36}
\partial_{2} \partial_{\overline{2}}\nu = 0. 
\end{equation}

\noindent {\bf 3.   Gauge invariance of the  Ansatz:}

Before considering explicit parametrizations of
${\cal M}^{(1)}_{2}$ and ${\cal M}^{(2)}_{2}$ based on
solutions of (\ref{31}) -- (\ref{36}), let us investigate
the effect of the $ U(1)$ transformation
\begin{equation}\label{37}
\psi_{1} \rightarrow  exp(i \alpha (x,y,s,t)) \psi_{1}
\end{equation}
\noindent on (\ref{15}), (\ref{16}), (\ref{17}). The
transformation (\ref{37}) is admitted by the RHS of
(\ref{17}), hence one may wonder whether it leads to new
and independent solutions  through (\ref{19}), (\ref{20})
and (\ref{23}).  The answer is that it does not; the
transformation
\begin{equation}\label{38}
A_{\mu} \rightarrow  A_{\mu} - \partial_{\mu} \alpha
\end{equation}
\noindent accompanying (\ref{37}) is forced to have the special form
\begin{equation}\label{39}
\alpha = \theta (x,y) + \lambda (s,t)
\end{equation}
\noindent by the SWME; hence the special form  of
$A_{\mu}$ in equation (\ref{4}) of the Ansatz is preserved.
This can be shown most easily by using the complex combinations
$\partial_{s} + i\partial_{t} =
2\partial_{\overline{2}},~~A_{s}+iA_{t}=
2A_{\overline{2}}$, etc.  After setting $\psi_{2}=0$
and applying (\ref{37}), equations (\ref{15}) and (\ref{16}) become
\begin{equation}\label{40}
(2 \partial_{\overline{2}}+\partial_{\overline{2}} \nu  +
2iA_{\overline{2}} + 2i\partial_{\overline{2}} \alpha) \psi_{1}=0
\end{equation}
\noindent  and
\begin{equation}\label{41}
(2 \partial_{\overline{1}}+\partial_{\overline{1}} \mu
+ 2iA_{\overline{1}} + 2i\partial_{\overline{1}} \alpha) \psi_{1}=0.
\end{equation}
\noindent Recalling that $\psi_1 = \zeta + i\eta$ and $\zeta $,
$\eta$ are both real constants, it follows that
\begin{equation}\label{42}
i \partial_{\overline{2}} \alpha = -iA_{\overline{2}}-
{1 \over 2}\partial_{\overline{2}} \nu
\end{equation}
\noindent and
\begin{equation}\label{43}
i \partial_{\overline{1}} \alpha =
-iA_{\overline{1}}-{1 \over 2}\partial_{\overline{1}} \mu ,
\end{equation}
\noindent which can only hold if $\alpha$ has the
form (\ref{39}) (recall that since $\alpha$ is real, we cannot add a
function holomorphic in $z_{1}$ and $z_{2}$ to it).  This is of course
because $A_{\overline{1}} =
A_{\overline{1}}(z_{1},\overline{z_{1}})$,
$\mu=\mu (z_{1},\overline{z_{1}})$ and $A_{\overline{2}} =
A_{\overline{2}}(z_{2},\overline{z_{2}})$,
$\nu=\nu (z_{2},\overline{z_{2}})$.  {\em Thus the Ansatz is
only compatible with gauge transformations of the form
(\ref{39}), which corresponds to having separate $U(1)$ fibers
over each of the two-manifolds}.  Interestingly, (\ref{42}) and
(\ref{43}) partly realize the original version of H. Weyl's
"Eichinvarianz" \cite{Weyl}, which was meant to fix the form of
the electromagnetic coupling by demanding that field equations
transform covariantly under local conformal transformations.

\noindent {\bf 4.  Solutions of the form $\Sigma_{p_{1}}
\times \Sigma_{p_{2}}$ :}

\noindent {\bf 4.1   Cases (ii) and (iii):} These are described by
the equations (\ref{33})--(\ref{36}).  Both represent the same solution
with one flat surface and one surface of constant negative curvature;
they can be transformed ino each other by
$\mu \leftrightarrow \nu, z_{1} \leftrightarrow  z_{2}$.
It will thus suffice to consider, say, only (\ref{33}) and
(\ref{34}), which we now do.

We may start by taking $\mu$ as the real part of an analytic
function $\zeta (z_{1})$.  The resulting metric for
${\cal M}^{(1)}_{2}$ is then of the form
\begin{equation}\label{44}
ds^{2}({\cal M}^{(1)}_{2})=\exp (\zeta (z_{1})+
\overline{\zeta} (\overline{z_{1}})) dz_{1} d\overline{z_{1}}
\end{equation}
\noindent which can be flattened by going over to the
coordinate $\tilde{z_{1}}(z_{1})= \int \exp \zeta (z_{1})  dz_{1}$.
${\cal M}^{(1)}_{2}$ can then be turned  into a flat torus by
choosing $\tilde{z_{1}}(z_{1})$ as an inverse elliptic function,
tesellating the  $\tilde{z_{1}}$ plane into parallelograms.

A similar  procedure  can  be effected on the one-sheeted
constant negative curvature hyperboloid stereographically
projected onto the  complex $g_{2}(z_{2})$ plane via
Liouville's \cite{Liou} solution to (\ref{34}), which reads
\begin{equation}\label{45}
\nu={1 \over 2} \ln \frac{4|\frac{dg_{2}}{dz_{2}}|^{2}}
{(1-g_{2}\overline g_{2})^{2}}
\end{equation}
\noindent At this point, $g(z_{2})$ is an arbitrary analytic
function.  We have therefore the familiar Kleinian metric for
${\cal M}^{(2)}_{2}$:
\begin{equation}\label{46}
ds^{2}({\cal M}^{(1)}_{2})=e^{2\nu} ~ dz_{2}~d\overline{z_{2}}=
\frac{dg_{2}d\overline{g_{2}}}{(1-g_{2}\overline{g_{2}})^{2}}
\end{equation}
\noindent One can of course also put the metric into the Poincar\'{e}
form
\begin{equation}\label{47}
ds^{2}({\cal M}^{(2)}_{2})=\frac{df_{2}d\overline{f_{2}}}
{(Im f_{2})^{2}}
\end{equation}
\noindent via $g_{2}(z_{2})=\frac{(f_{2}-i)}{(f_{2}+i)}$ which
maps the interior of the circle $|g_{2}(z_{2})|^{2}=1$ to the
upper-half $f$-plane $C_{+}$.  If one now chooses $f_{2}(z_{2})$
as the Fuchsian function used in uniformizing an algebraic
function whose Riemann surface has genus $p_{2}$
\cite{Ford}, $C_{+}$ gets tesellated into $4p_{2}$-gons with
geodesic edges and the manifold ${\cal M}^{(2)}_{2}$ becomes
the Riemann surface $\Sigma_{p_{2}}$ of genus $p_{2}$ when the
edges are identified in the standard way described
in, say, \cite{Dubr}.  For explicit parametrizations of
Fuchsian functions we refer the reader to \cite{Nehari}.

To summarize, the 4-manifolds for cases (ii) and (iii)
are of the form $(\Sigma_{1}\equiv T^{2})\times \Sigma_{p}, p \geq 2$.
Setting the first two components of $A_{\mu}$ which are pure gauge
equal to zero, the solution for case (ii) becomes
\begin{equation}\label{48}
A= \frac{i}{2} \{ \frac{1}{2}d\ln (\frac{d\overline{g}_{2}}
{d\overline{z}_{2}}
\frac {dz_{2}}{dg_{2}}) + 
\frac {(g_{2}d\overline{g}_{2}- \overline{g}_{2}dg_{2})}
{(1- g_{2}\overline{g}_{2})} \},
\end{equation}
\begin{equation}\label{49}
\omega^{34}= - 2A,
\end{equation}
\begin{equation}\label{50}
F= - \frac{1}{2}R^{34}=  i\frac{dg_{2} \wedge 
d \overline{g}_{2}}{(1 - g_{2}\overline{g}_{2})^2},
\end{equation}
\begin{equation}\label{51}
\Psi ^{T}= (\psi_{1},0,0,0), \psi_{1}=cst.,~~R=R^{ab}_{ab}=
-2|\psi_{1}|^{2}.
\end{equation}
\noindent In order to obtain the corresponding formulae for
case (iii) we simply switch the set $(\mu=0, \nu=\nu (z_{2},
\overline{z_{2}}), g_{2}(z_{2}), \omega ^{3}_{4}, R^{3}_{4})$
with $(\nu=0, \mu=\mu (z_{1}, \overline{z_{1}}), g_{1}(z_{1}),
\omega ^{1}_{2}, R^{1}_{2})$.

\noindent {\bf 4.2   Case (i):}

The pair of Liouville equations (\ref{31}) and (\ref{32})
(with the upper sign on the RHS) correspond to the 4-manifold
$\Sigma_{p_{1}} \times \Sigma_{p_{2}}~(p_{1}, p_{2} \geq 2)$,
where the scalar curvature of the Riemann surfaces
are $-2|\phi|^{2}$ and $-2(|\psi_{1}|^{2} - |\phi|^{2})$,
respectively.  The connections and the curvatures are
then given by
\begin{equation}\label{52}
\omega^{1}_{2}= - i \{ \frac{1}{2}d\ln (\frac{d\overline{g}_{1}}
{d\overline{z}_{1}}
\frac {dz_{1}}{dg_{1}}) + 
\frac {(g_{1}d\overline{g}_{1}- \overline{g}_{1}dg_{1})}
{(1- g_{1}\overline{g}_{1})} \},
\end{equation}
\begin{equation}\label{53}
\omega^{3}_{4} = - i \{ \frac{1}{2}d\ln (\frac{d\overline{g}_{2}}
{d\overline{z}_{2}}
\frac {dz_{2}}{dg_{2}}) + 
\frac {(g_{2}d\overline{g}_{2}- \overline{g}_{2}dg_{2})}
{(1- g_{2}\overline{g}_{2})} \},
\end{equation}
\begin{equation}\label{54}
A = - \frac{1}{2}(\omega^{1}_{2}+\omega^{3}_{4}), 
\end{equation}
\begin{equation}\label{55}
R^{1}_{2} = - 2i \frac {dg_{1} \wedge d\overline{g}_{1}}
{(1- g_{1}\overline{g}_{1})^{2}},~~~R^{3}_{4} = - 2i \frac {dg_{2}
\wedge d\overline{g}_{2}} {(1- g_{2}\overline{g}_{2})^{2}},
\end{equation}
\begin{equation}\label{56}
F= i \frac {dg_{1} \wedge d\overline{g}_{1}}
{(1- g_{1}\overline{g}_{1})^{2}}+
i \frac {dg_{2} \wedge d\overline{g}_{2}}
{(1- g_{2}\overline{g}_{2})^{2}}= - \frac{1}{2}(R^{1}_{2} +R^{3}_{4}),
\end{equation}

\noindent {\bf 4.3   Case (iv):}

This is based on (\ref{31}) and the lower (negative) sign on
the RHS of (\ref{32}).  The local form of the solution
of (\ref{32}) is then
\begin{equation}\label{57}
\nu = \frac{1}{2} \ln \frac{4|\frac{dg_{2}}{dz_{2}}|^{2}}
{(1+ g_{2}\overline{g}_{2})^{2}},
\end{equation}
\noindent  which means ${\cal M}^{(2)}_{2} = S^{2}$ with radius 
$(|\phi|^{2}| -|\psi_{1}|^{2})^{-1/2}$ in the original
dimensionless coordinates.  As in the other cases,
${\cal M}^{(1)}_{2}$ is a surface of constant negative
curvature, from which a Riemann surface $\Sigma_{p_{2}}
\geq 2$ can again be constructed via tesellation by $4p_{2}$-gons.
\begin{equation}\label{58}
\omega^{1}_{2}= - i \{ \frac{1}{2}d\ln (\frac{d\overline{g}_{1}}
{d\overline{z}_{1}}
\frac {dz_{1}}{dg_{1}}) + 
\frac {(g_{1}d\overline{g}_{1} - \overline{g}_{1}dg_{1})}
{(1- g_{1}\overline{g}_{1})} \},
\end{equation}
\begin{equation}\label{59}
\omega^{3}_{4} = i \{ \frac{1}{2}d\ln (\frac{d\overline{g}_{2}}
{d\overline{z}_{2}}
\frac {dz_{2}}{dg_{2}}) + 
\frac {(g_{2}d\overline{g}_{2}- \overline{g}_{2}dg_{2})}
{(1+ g_{2}\overline{g}_{2})} \},
\end{equation}
\begin{equation}\label{60}
A = - \frac{1}{2}(\omega^{1}_{2}+\omega^{3}_{4}), 
\end{equation}
\begin{equation}\label{61}
R^{1}_{2} = - 2i \frac {dg_{1} \wedge d\overline{g}_{1}}
{(1- g_{1}\overline{g}_{1})^{2}},~~~R^{3}_{4} =
+2i \frac {dg_{2} \wedge d\overline{g}_{2}}
{(1+ g_{2}\overline{g}_{2})^{2}},
\end{equation}
\begin{equation}\label{62}
F= i \frac {dg_{1} \wedge d\overline{g}_{1}}
{(1- g_{1}\overline{g}_{1})^{2}}
- i \frac {dg_{2} \wedge d\overline{g}_{2}}
{(1+ g_{2}\overline{g}_{2})^{2}}= - \frac{1}{2}(R^{1}_{2} +R^{3}_{4})~.
\end{equation}
\noindent In spite of the apparent similarity between the solutions
of this section and the previous one, there is one very important
difference:  since ${\cal M}^{(2)}_{2} = S^{2}$, $g_{2}(z_{2})$
is not a Fuchsian function; in fact it can only be
\begin{equation}\label{63}
g_{2}(z_{2}) =  \frac{az_{2} + b}{cz_{2} + d}, 
\end{equation}
\noindent which represents the most general one-to-one mappings
of  $S^{2}$ to itself.
Finally, it is easy to show that the solutions in all the cases
are entirely invariant under transformations of the form
\begin{equation}\label{64}
\tilde{g_{i}} =  \frac{\alpha g_{i} + \beta }
{\pm \overline{\beta}g_{i} + \overline{\alpha}},~~~\alpha
\overline{\alpha} \mp +  \beta \overline{\beta},~~i=1, 2,
\end{equation}
\noindent where the upper (lower) sign applies in the negative
(positive) curvature cases.  By this we mean that not only curvatures
but even the connection 1-forms $\omega ^{ab}$ and $A$ remain
unchanged.  Thus the case (i) is invariant under
$SU(1,1)\times SU(1,1)$, where both $g_{1}(z_{1})$
and  $g_{2}(z_{2})$ are subjected to (\ref{64}).  The
cases (ii)--(iii) and (iv) have invariance groups $SU(1,1)$
and $SU(1,1)\times SU(2)$, respectively.  When one passes
from the Klein form (\ref{40}) to the Poincar\'{e}
form (\ref{41}) of the hyperboloid metric,
(\ref{64}) (upper sign) is replaced by
\begin{equation}\label{65}
f_{i} = \frac{af_{i} + b}{cf_{i} + d}~, ~~ab - cd =1,
\end{equation}
where $a, b, c, d$ are all real numbers.  This is of course 
$SL(2, \bf R)$ rather than $SU(1,1)$.  The $SL(2, \bf Z)$
subgroup of $SL(2, \bf R)$ shuffles $4p$-gons of the tesellation
amongst themselves; invariance under this integer subgroup
ensures that one has the same solution in each domain.

\noindent {\bf 5.   Topological numbers and the uniqueness of the solution:}

For a 4-dimensional manifold  ${\cal M}_{4}$, the signature
$\sigma$ is given by \cite{Egu}
\begin{equation}\label{66}
\sigma ({\cal M}_{4}) = - \frac{1}{24 \pi^{2}}
\int_{{\cal M}_{4}} R^{a}_{b} \wedge R^{b}_{a}.
\end{equation}
\noindent For the solutions considered here, the only non-vanishing
components of $R^{a}_{bcd}$ are $R^{1}_{212}$ and $R^{3}_{434}$;
hence $\sigma$ trivially vanishes in all the cases (i)--(iv).

The Euler characteristic is most easily computed from the Kunneth formula  
\begin{equation}\label{67}
\chi ({\cal M}_{4}) = \chi ({\cal M}^{(1)}_{2})
\chi ({\cal M}^{(2)}_{2})=(2 - 2 p_{1})(2 - 2p_{2}).
\end{equation}
Finally, we must evaluate 
\begin{equation}\label{68}
c_{1}^{2} =\frac{1}{(2\pi)^{2}}\int_{{\cal M}_{4}} F^{2},
\end{equation}
\noindent where, as in (\ref{56}) and  (\ref{62}), we can write
\begin{equation}\label{69}
F=  - \frac{1}{2}(R^{1}_{2} +R^{3}_{4})
\end{equation}
and then use (\ref{67}) and the Gauss-Bonnet theorem to obtain 
\begin{equation}\label{70}
c_{1}^{2} = \frac{1}{2} \chi ({\cal M}_{4}).
\end{equation}
\noindent Then, using the notation and conventions of
\cite{Wit1}, the virtual dimension $W$ of the moduli space defined by
\begin{equation}\label{71}
W = - \frac{(2\chi + 3 \sigma)}{4} + c_{1}^{2} = 0 
\end{equation}

\noindent is seen to vanish.   Thus the solutions of the type $\Sigma_{p_{1}} \times \Sigma_{p_{2}} $ are unique up to $U(1)$ transformations in the bundle and conformal changes  of the metrics.  

\noindent {\bf 6.  A self-dual solution of the Einstein-Maxwell-Dirac
equations:}

The similarities between the solutions of the SWME considered  so
far and the Bertotti-Robinson solution (covariantly constant
electromagnetic fields on a product of two two-manifolds) suggest
that some of the field configurations found so far may satisfy the
Einstein-Maxwell-Dirac field equations  
\begin{equation}\label{72}
{\cal R}_{\mu\nu}-\frac{1}{2}g_{\mu\nu}{\cal R}=
\kappa(T_{\mu\nu}(\mbox{\small e.m.})+
T_{\mu\nu}(\mbox{\small Dirac}))+\Lambda g_{\mu\nu},
\end{equation}
\begin{equation}\label{73}
F^{\mu\nu}_{;\mu}= j^{\nu}_{\small el.}= 
\psi^{\dagger} \gamma^{\nu} \psi,
\end{equation}
\begin{equation}\label{74}
\tilde{F}^{\mu\nu}_{;\mu}=k^{\nu}_{\small mag.}= 
\psi^{\dagger} \gamma^{\nu} \gamma ^{5} \psi.
\end{equation}
\noindent The Dirac equation, already satisfied by our
solutions, (\ref{1}) completes the system of coupled equations
our three interacting fields have to obey.  Using a Weyl spinor in
Euclidean signature (which is why the usual $\bar{\psi}$'s are
replaced by $\psi^{\dagger}$'s) immediately gives
\begin{equation}\label{75}
\psi^{\dagger}\gamma^{\nu}\psi=\psi^{\dagger}\gamma^{\nu}\gamma ^{5}\psi = 0
\end{equation}
\noindent and
\begin{equation}\label{76}
T_{\mu\nu}(\mbox{\small Dirac}) = 
\frac{i}{2}[\psi^{\dagger} \gamma^{\mu} D_{\nu}\psi    -
(D_{\nu}\psi^{\dagger}) \gamma^{\mu} \psi]= 0
\end{equation}
\noindent because the $\gamma_{\mu}$ are block off-diagonal
while the covariant derivative is block diagonal.  Taking the trace
of both sides of (\ref{72}) yields
\begin{equation}\label{77}
{\cal R}= - 4\Lambda,
\end{equation}
\noindent or, using (\ref{27}),
\begin{equation}\label{78}
\Lambda = \frac{|\psi_{1}|^{2}}{2}= constant,
\end{equation}
\noindent showing that the masless monopole condensate provides
a cosmological constant.

Now we know that $F_{\mu \nu}$ is covariantly constant for  all our
SWME solutions, i.e.,
\begin{equation}\label{79}
D^{\mu}F_{\mu \nu} = D^{\mu} \tilde{F}_{\mu \nu} = 0
\end{equation}
\noindent and that these are consistent with (\ref{73}),
(\ref{74}) and (\ref{75}); we however must still satisfy the by
now considerably simplified Einstein field equations
\begin{equation}\label{80}
{\cal R}_{\mu\nu}-\frac{1}{2}g_{\mu\nu}{\cal R}=
\kappa T_{\mu\nu}(\mbox{\small e.m.})+
\Lambda g_{\mu\nu}.
\end{equation}  
\noindent This is most easily accomplished by imposing
\begin{equation}\label{81}
F_{\mu \nu} =  \tilde{F}_{\mu \nu} =
\frac{1}{2} \epsilon_{\mu \nu \alpha \beta}F^{\alpha \beta}
\end{equation}
\noindent which can be seen to hold when
\begin{equation}\label{82}
|\phi|^{2} = \frac{1}{2}|\psi|^{2}
\end{equation}
\noindent by going back to (\ref{22}), (\ref{25}), (\ref{28})
and (\ref{29}).  This means not only $F_{\mu \nu} $ but the
whole solution is self-dual in the sense that the manifolds
${\cal M}^{(1)}_{2}$ and
${\cal M}^{(2)}_{2}$ are identical!  Furthermore, explicit
calculation, both analytical and by using REDUCE, shows that
none of our other SWME solutions obeys (\ref{72}) -  (\ref{74}).

\noindent {\bf 7.   K\"{a}hler structure and relation to the
Monge-Amp\`{e}re equations:}

In the complex coordinates $z^{a} (a=1,2)$ the metric
$g_{a\bar{b}}$ of ${\cal M}_{4}$ can be obtained from the
K\"{a}hler potential
\begin{equation}\label{83}
U = \mu + \nu = \frac{1}{2} 
\ln \frac{{g'}_{1} {\overline{g'}}_{1}{g'}_{2}
{\overline{g'}}_{2}}{(1- g_{1}\overline{g}_{1})^
{2}(1- g_{2}\overline{g}_{2})^{2}}
\end{equation}
\noindent through the K\"{a}hler form
\begin{equation}\label{84}
K = \frac{i}{2}g_{a\bar{b}}dz^{a}\wedge d\overline{z}^{b} 
= \frac{i}{2}\partial \overline{\partial} U. 
\end{equation}
Now it is known \cite{Egu} that if a  K\"{a}hler potential
$\tilde{U}$ satisfies the Monge-Amp\`{e}re equations
\begin{equation}\label{85}
(\overline{\partial}\partial\tilde{U})
\wedge(\overline{\partial}\partial \tilde{U})=
\exp{2\tilde{\Lambda}\tilde{U}}~d\overline{z}^{1}\wedge  dz^{1}
\wedge d\overline{z}^{2}\wedge dz^{2},
\end{equation}
\noindent then the metric for $\tilde U$ defined via (\ref{84})
solves Einstein's equations with cosmological constant
$\tilde{\Lambda}$; in particular, the Weyl tensor is self-dual.
The metric in (\ref{84}) can be related to (\ref{85}) by choosing
\begin{equation}\label{86}
\tilde{U} = \frac{1}{\tilde{2\Lambda}}(2U - 3 \ln 2).
\end{equation}
\noindent  Note that $\tilde{\Lambda}$  is dimensionless in the
dimensionless coordinates used in (\ref{83}-\ref{86}); it can of
course always be scaled back together with the coordinates to
the value $\Lambda = \frac{|\psi_{1}|^{2}}{2}$ in (\ref{78}).
The Liouville equations (\ref{31}) and (\ref{32}) then
guarantee (\ref{85}).  The Weyl tensor is self-dual in
the trivial sense that all its components vanish.
\noindent {\bf  8.  Concluding remarks:}

One might wonder what would have happened had we chosen the
other component of the Weyl spinor to vanish in (\ref{15})
and (\ref{16}).   Going through the formulae, it is not difficult
to see that complex variables are replaced by their conjugates,
leading to a sign change in the electromagnetic potential, while
the metric remains unchanged. Recalling that two components of
the Weyl spinor correspond to particles and antiparticles, we
see that the new solution is just the charge conjugate of the
original one.  We now have an antimonopole condensate and the
sense of the vortices is reversed.

It is worth emphasizing here some surprisingly general facts
about Weyl spinors and $U(1)$ connections on a 4-manifold of
Euclidean signature.  As noted in section 6, (\ref{75})
and (\ref{76}) automatically hold.  If, in addition, one
imposes self-duality on the $F_{\mu \nu}$, Einstein's
equations are immediately reduced to the defining equation
for an Einstein space and, although not explicitly stated
by LeBrun, this is what allows one to consider Einstein spaces
in a SW setting without having to take Maxwell's equations and
matter tensors into account.  Since there is nothing else
available, the SW equations then force an identification of
the cosmological constant with the monopole condensate.  The
physical picture is astonishing, at least to the author, in
that Weyl spinors and self-dual $U(1)$ gauge fields can
seemingly live on an Einstein manifold of Euclidean signature
without having any other effect on the metric whatsoever!
Interestingly, this also "saturates" the LHS of (\ref{2})
by forcing the curvature to be purely self-dual. 

\noindent {\bf Acknowledgements}

I am grateful to S. Akbulut for answering many questions about
Seiberg-Witten theory, to P. Argyres for clarifying a point
involving the virtual dimension, and to Y. Nutku for pointing
out the connections to the Bertotti-Robinson solution and the
Monge-Amp\`{e}re equations, as well as carrying out the REDUCE
check.  I thank M. Ar\i k, T. Dereli and R. G\"{u}ven for useful
discussions and S. Nergiz for help with LATEX.

\end{document}